\documentclass[a4paper,10pt,twoside]{cpc-hepnp}
\usepackage{amsmath}
\usepackage{multicol}
\usepackage{graphicx}
\usepackage{booktabs}
\usepackage{amssymb,bm,mathrsfs,bbm,amscd}
\usepackage{lastpage}
\usepackage{lineno}
\usepackage{color}

\begin{document}

\title{The effect of heavy metal in CMOS on neutron induced single event upset simulated with Geant4\thanks{This work was supported by the National Natural Science Foundation of China (Grant No. 11235008) and partly supported by (Grant No. 10979010)}}

\author{%
      Huan ZHANG$^{1,2}$%
\quad Si-guang WANG$^{1;1)}$\email{siguang@pku.edu.cn}
\quad Wei CHEN$^{2}$
\quad Shan-chao YANG$^{2}$
}

\maketitle

\address{%
$^1$ School of Physics and State Key Laboratory of Nuclear Physics and Technology, Peking University, Beijing 100871, China\\
$^2$ Northwest Institute of Nuclear Techniques, Xi'an 710024, China\\
}

\begin{abstract}
Local metal interconnection is widely used in modern complementary metal oxide semiconductor (CMOS) technology. The most frequently used local materials are some heavy metals, such as tungsten (W) or copper (Cu). It's well known that single event upset (SEU) could occur in a CMOS under neutron exposure. In this paper the rectangular parallelepiped (RPP) method is used to investigate the SEU response of a typical CMOS. SEU induced by 1$\sim$14\:MeV neutrons are simulated with Geant4 and the cross sections are calculated. The results show that only in the structure with W, secondary particle $\alpha$ is created and SEUs are generated when the energy of neutron is less than $4$\:MeV.
\end{abstract}



\begin{multicols}{2}

\section{Introduction}

Since some heavy metals, such as copper and tungsten, are better conductors than aluminium, metal components for interconnection of semiconductor chips can be smaller in size with them, and consume less energy when electricity passes through them while keeping higher-performance. Therefore, in semiconductor integrated circuits, tungsten and copper are widely used.

In complementary metal oxide semiconductor (CMOS), incident neutrons, protons, and heavy-ions can induce single event upset (SEU) in its sensitive volume (SV). This paper only focus on neutron induced single event upset. In silicon, secondary particles created by nuclear interactions of incident neutron can range from H to Si, eventually P, including isotopes~\citep{lab1}. According to the primary neutron energy and the details of the interaction, all these particles have variable energies, linear energy transfers (LETs) and free paths. SEU is caused by the electrical changes associated to the electron-hole pairs generated along the ionizing particle track. When the number of electron-hole pairs is more than the critical charge in the SV of the CMOS, an SEU will occur.

The introduce of heavy metal in CMOS may affect the SEU rates of neutron. To study the effect of heavy metal quantitatively, we use tungsten interconnection as an example. In this paper, a rectangular parallelepiped model of a CMOS is built, then the SEU rates induced by 1$\sim$14\:MeV neutrons are simulated with Geant4~\citep{g4}. The results of SEU cross sections with and without tungsten are compared and the genesis of the difference is explored.

\section{Method}

To simplify the structure, the rectangular parallelepiped (RPP) method~\citep{lab2} is chosen to do the simulation in this work. The method, which simplified the structure by using a collection of rectangular parallelepiped, is much easier to model the structure of CMOS with Geant4 than a detail description. Deposited energy is used to determine an upset instead of total charge in the SV because the number of electron-hole pairs is too large to simulate. An upset will occur if and only if the deposited energy is more than its critical value ($\text{E}_c$), corresponding to the critical charge ($\text{Q}_c$).

Two multi-layer planar targets are used to investigate the SEU response of a typical CMOS with a multi-layer metallization system, as shown in Fig.~\ref{fig1}. The sensitive volume in the silicon for these structures is a $3.16 \times 3.16 \times 2$\:$\mu m^{3}$ rectangular region which is located beneath the metallization stack, where the created electron-hole pairs can cause SEU.

There is only one layer has difference between the two structures. In one structure, this layer is composed of tungsten, as shown in (a) of Fig.~\ref{fig1}, which is commonly used in integrated circuits to provide electrical connections between layers of metallization or in contact with the underlying silicon~\citep{lab3}, while in the other structure it is silicon dioxide, referred to (b) of the figure. This layer is approximately $1.4$\:$\mu m$ above the sensitive volume with $0.6$\:$\mu m$ in thickness. Characters of other layers are also shown in Fig.~\ref{fig1}. Effects of the tungsten layer on the SEU cross section are presented in this paper.
\end{multicols}

\begin{center}
\includegraphics[width=16cm]{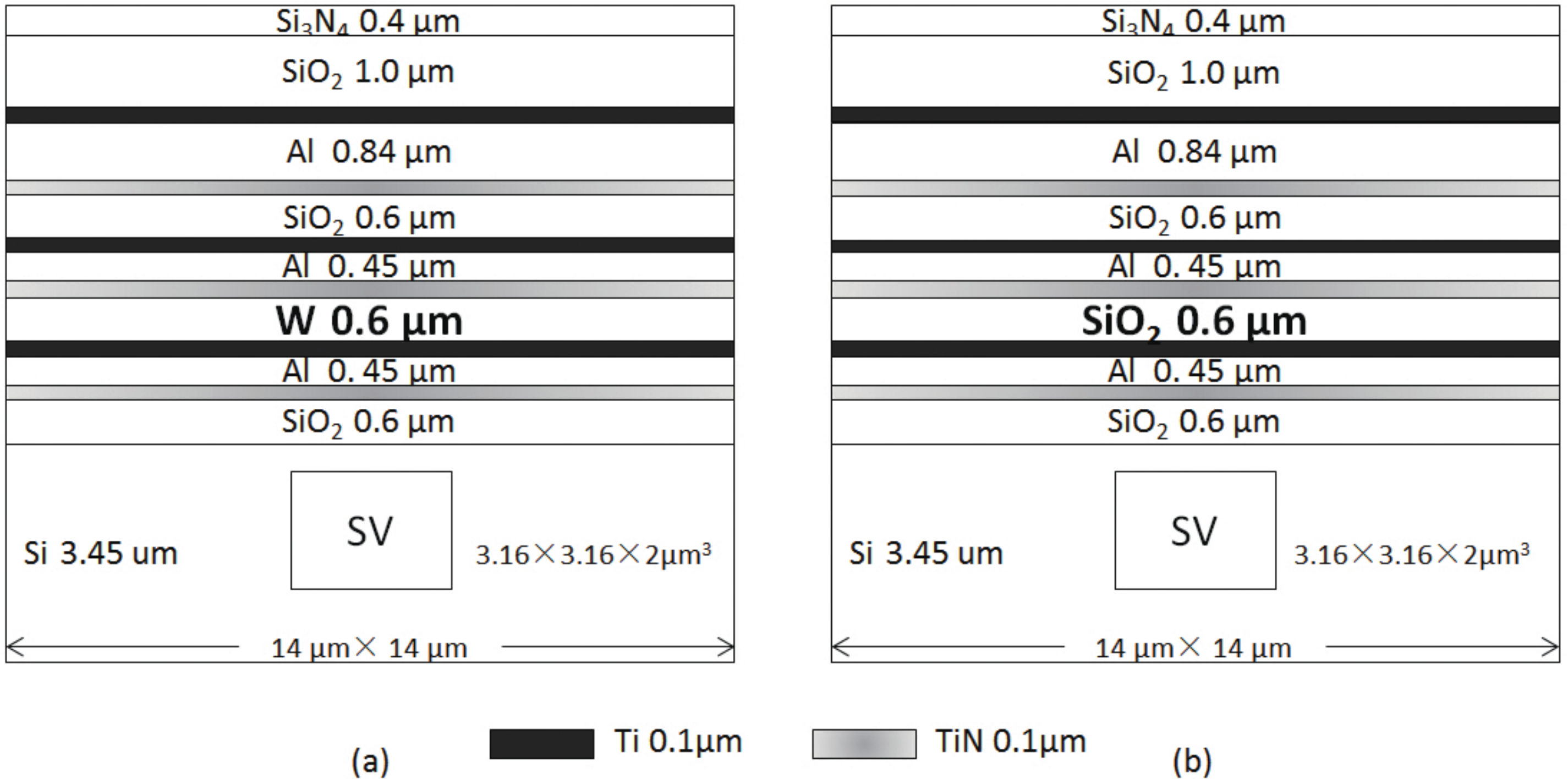}
\figcaption{\label{fig1}  Structures of the scaled CMOS without (a) and with (b) a tungsten layer. Only one layer has difference between the two plots.}
\end{center}

\begin{multicols}{2}

Energy deposition in the SV of the two structures is simulated with different mono-energetic beam of neutron. Total $6\times10^8$ neutrons are generated and perpendicularly inject into the top of the CMOS. The elastic scattering, inelastic scattering, neutron capture process and ionization process for the secondary ions are taken into account in our code.

In silicon, an electron-hole pair creation normally needs around $3.6$\:eV deposited energy~\citep{ehp}, so the relationship between $E_c$ and $Q_c$ is
\begin{displaymath}
 E_{c}(MeV) = 22.5 \times Q_{c}(pC).
\end{displaymath}
The critical energy of our CMOS is set as $0.69$\:MeV~\citep{cri}.

The cross section of single event upset can be obtained with the following relation:
\begin{displaymath}
 \sigma (E_{c},E_{n}) = \frac{ n_{SEU}}{\phi},
\end{displaymath}
where E$_{n}$ is the energy of incident neutron, $n_{SEU}$ is the number of SEU events with $E_{dep}>E_{c}$, $E_{dep}$ is deposited energy in the SV. And $\phi$ is the fluence of the beam, which is calculated with
\begin{displaymath}
\phi = \frac{N}{A},
\end{displaymath}
where N is the total number of incident neutron simulated and A is the irradiated area. It's worth to note that the cross section depends on $E_{n}$ and $E_{c}$, which is target geometry dependence.

\section{Results}
1$\sim$14\:MeV neutrons are simulated with Geant4 in this paper and the cross sections are calculated, respectively.
The simulation results of the two structures are shown in Fig.~\ref{fig2}.
\begin{center}
\includegraphics[width=8cm]{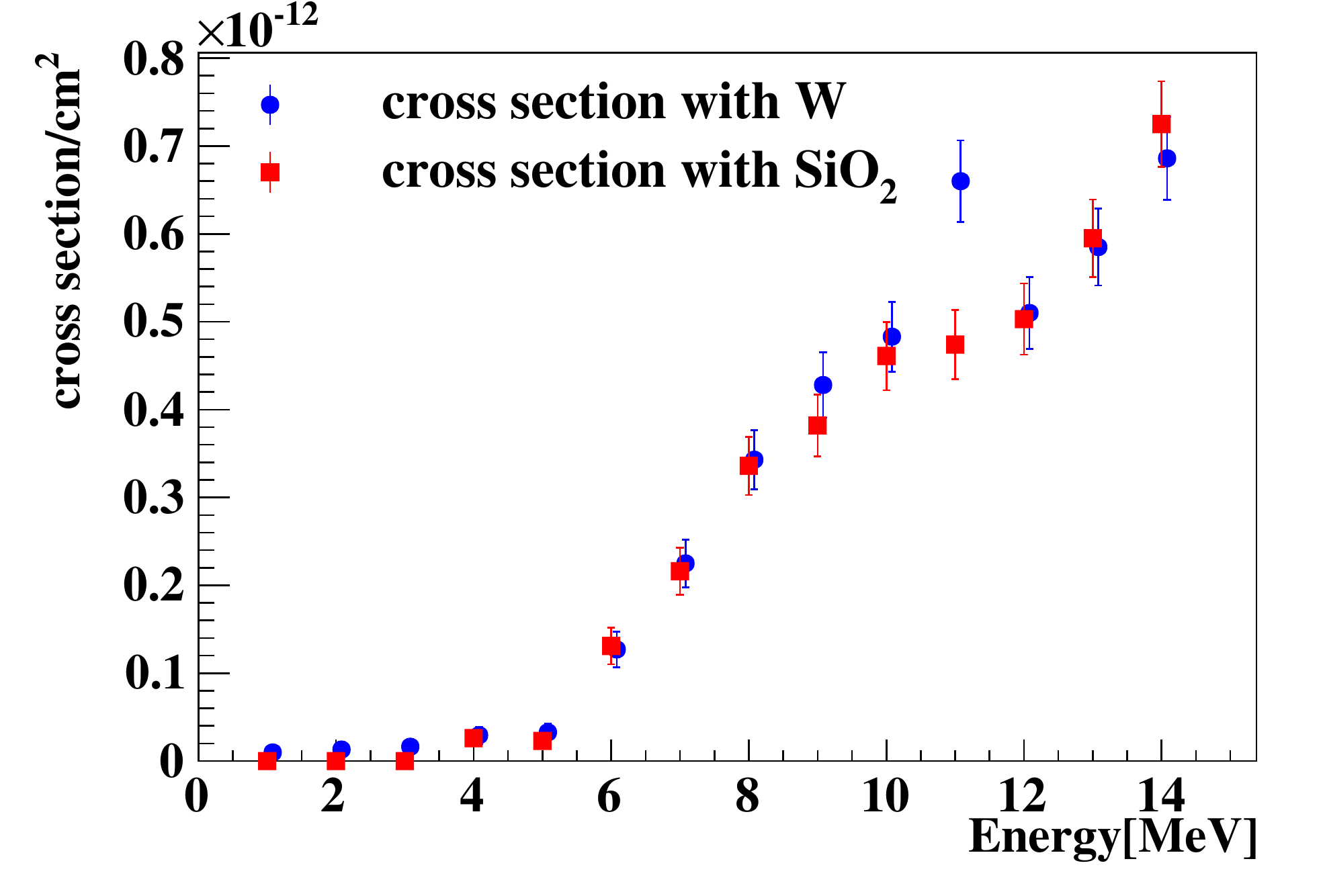}

(a)

\includegraphics[width=8cm]{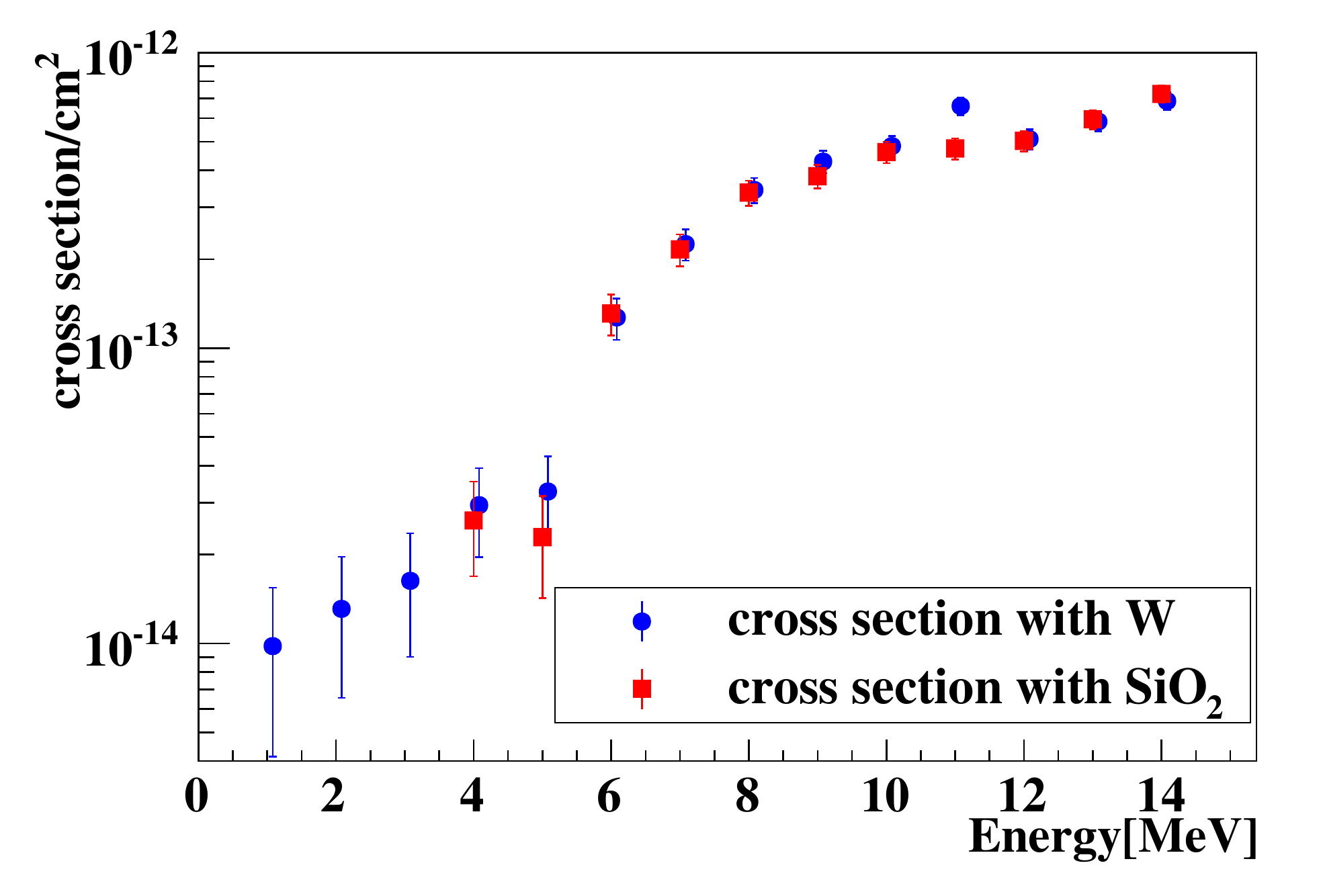}

(b)
\figcaption{\label{fig2}  SEU cross sections of the two structures simulated with different energy neutron. The blue filled circles denote the cross section of the structure with W, while the red filled rectangles for the one without W. The data in (b) are as same as (a), except shown with log-Y axis. The points are shifted slightly from their values to avoid overlapping.}
\end{center}


As shown in Fig.~\ref{fig2}, the SEU cross sections of the two structures are almost the same when the energy of neutron is higher than 4\:MeV. In other words, the tungsten layer almost has no effect on the SEU rates for high energy neutron ($>$ 4\:MeV). But when the energy of neutron is less than 4\:MeV, no SEU has been seen in the structure without a tungsten layer, indicating the tungsten layer enhanced the SEU probability in low energy area of neutron ($<$ 4\:MeV).

\begin{center}
\includegraphics[width=8cm]{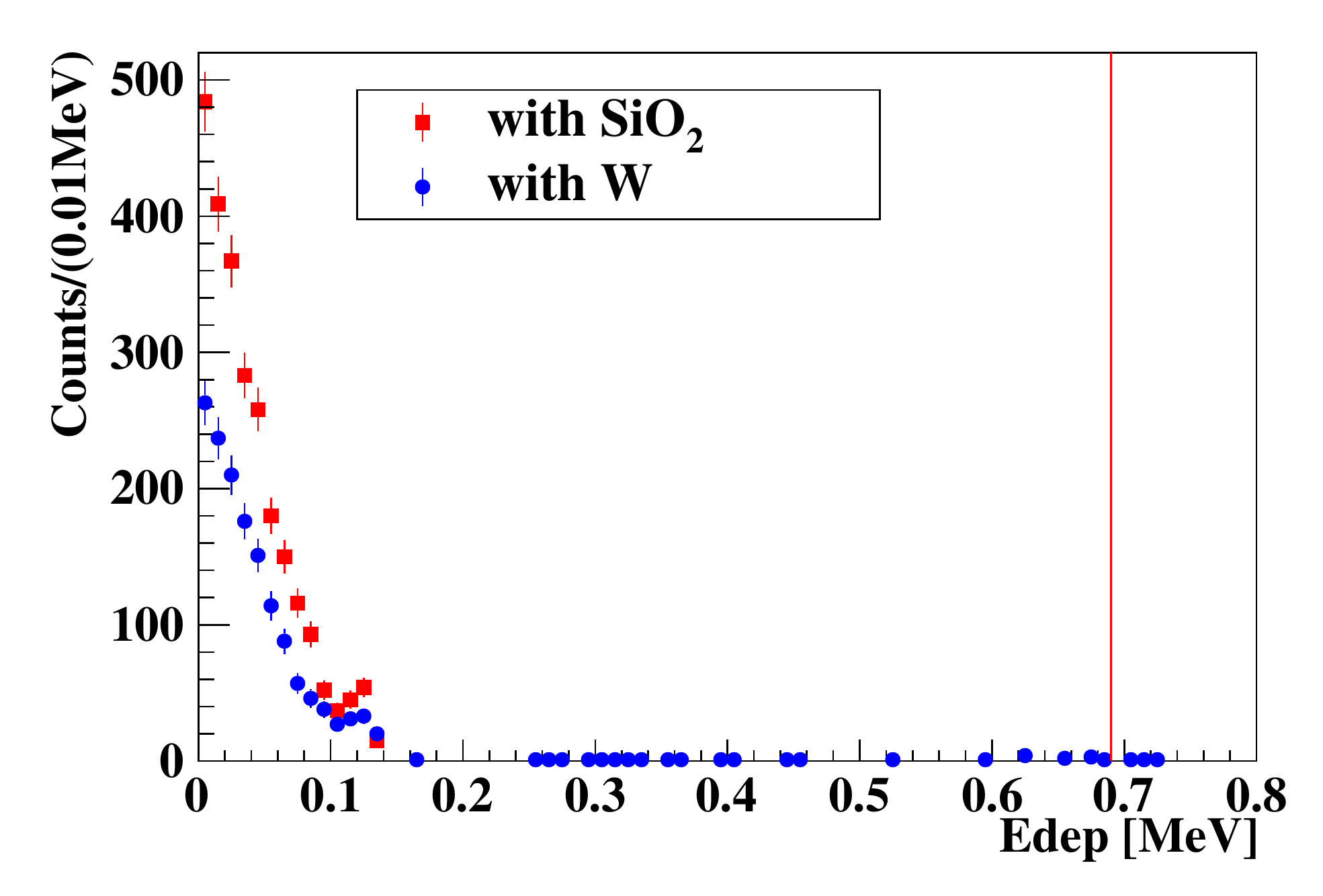}
\figcaption{\label{fig3}  The distribution of deposited energies in the SV. The blue filled circles denote the results for the structure with W, while the red filled rectangles for the one without W. Total $6\times10^8$ 1-MeV neutrons are simulated. The vertical red line is for $E_{c}=0.69$\:MeV.}
\end{center}


To explore the increase of SEU rates in the structure with W, the energy deposition distributions of 1\:MeV neutron in the structures with and without W are checked and shown in Fig.~\ref{fig3}. In the structure without a tungsten layer, no event has been seen with the deposited energy in the sensitive volume greater than the $E_{c}$, while there are 3 out of $6\times10^8$ events with deposited energies greater than the $E_{c}$ value in our simulation. The dominated contribution of energy deposition in the SV is from secondary particles created by neutrons through nuclear interactions. The total deposited energy of each kind of secondary particle in the SV from the simulation result is obtained and filled in a histogram as shown in Fig.~\ref{fig4}, which is normalized into one neutron. The difference between the two series of data indicates that the secondary particles in the two structures are different because of the existence of the tungsten layer.

\begin{center}
\includegraphics[width=8cm]{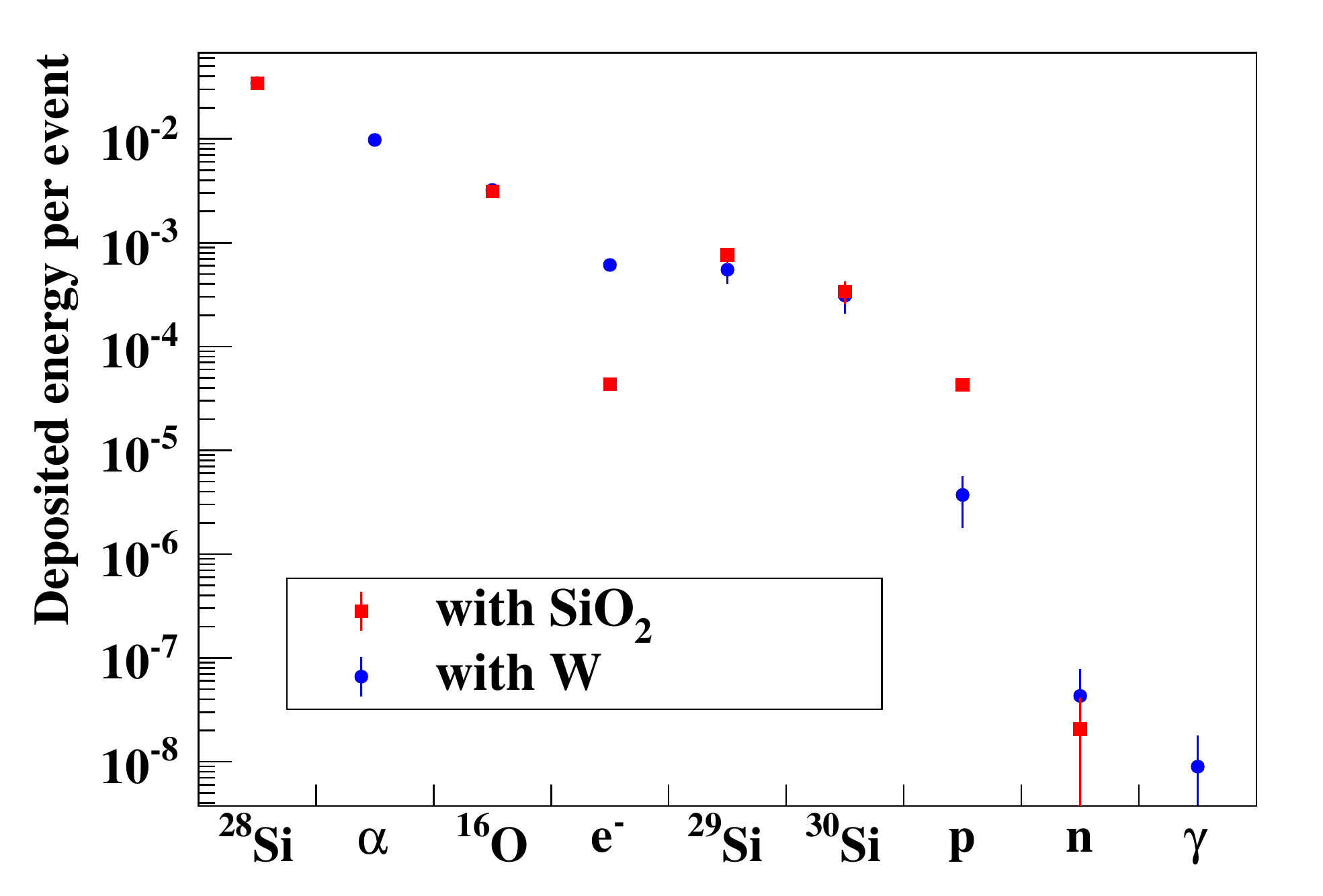}
\figcaption{\label{fig4}  Species of secondary particles and the corresponding energy deposition normalized into one neutron in the SV. The blue filled circles denote the results for the structure with W, while the red filled rectangles for the one with SiO2. Total $6\times10^8$ 1-MeV neutrons are simulated. No $\alpha$ and $\gamma$ has been seen in SV of the structure with SiO2. }
\end{center}

Figure~\ref{fig4} shows that secondary particles produced in both structures are $^{28}\text{Si}$, $^{29}\text{Si}$, $^{30}\text{Si}$, $^{16}\text{O}$, electron, and proton. The particle which produced the maximum total deposited energy is $^{28}\text{Si}$ in both structures. The distribution of deposited energy of $^{28}\text{Si}$ is shown in Fig.~\ref{fig5}, and that of $^{29}\text{Si}$, $^{30}\text{Si}$ and $^{16}\text{O}$ are shown in Fig.~\ref{fig6}. Plots for deposited energies from electrons, protons and $\gamma$s are ignored in this work since they are too few.

\begin{center}
\includegraphics[width=8cm]{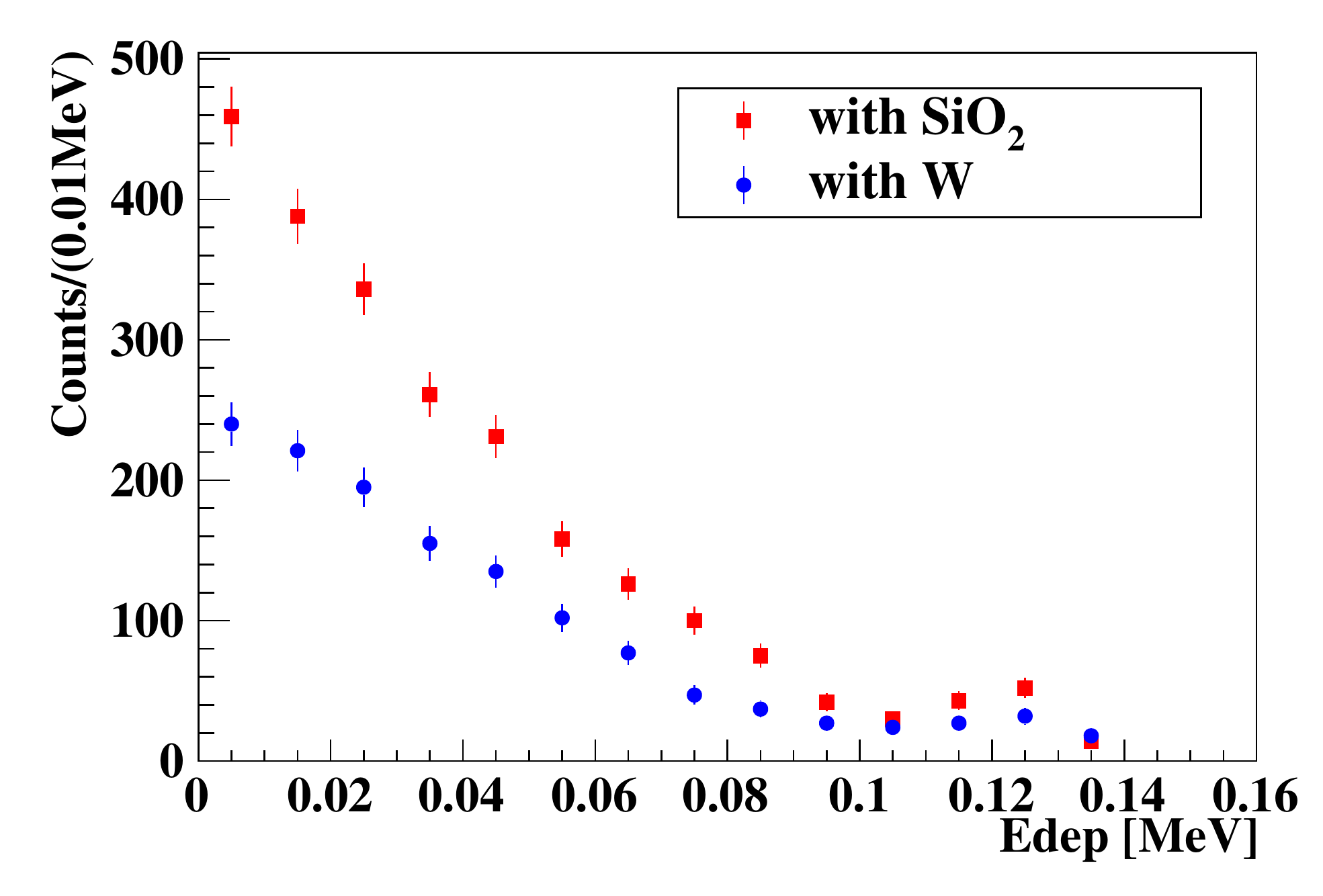}
\figcaption{\label{fig5}  Deposited energies of $^{28}\text{Si}$ in structures with and without tungsten layer. Total $6\times10^8$ 1-MeV neutrons are simulated. The blue filled circles denote the deposited energy for the structure with W, while the red filled rectangles for the one without W. Detailed description can be referred in the text.}
\end{center}

\begin{center}
\includegraphics[width=8cm]{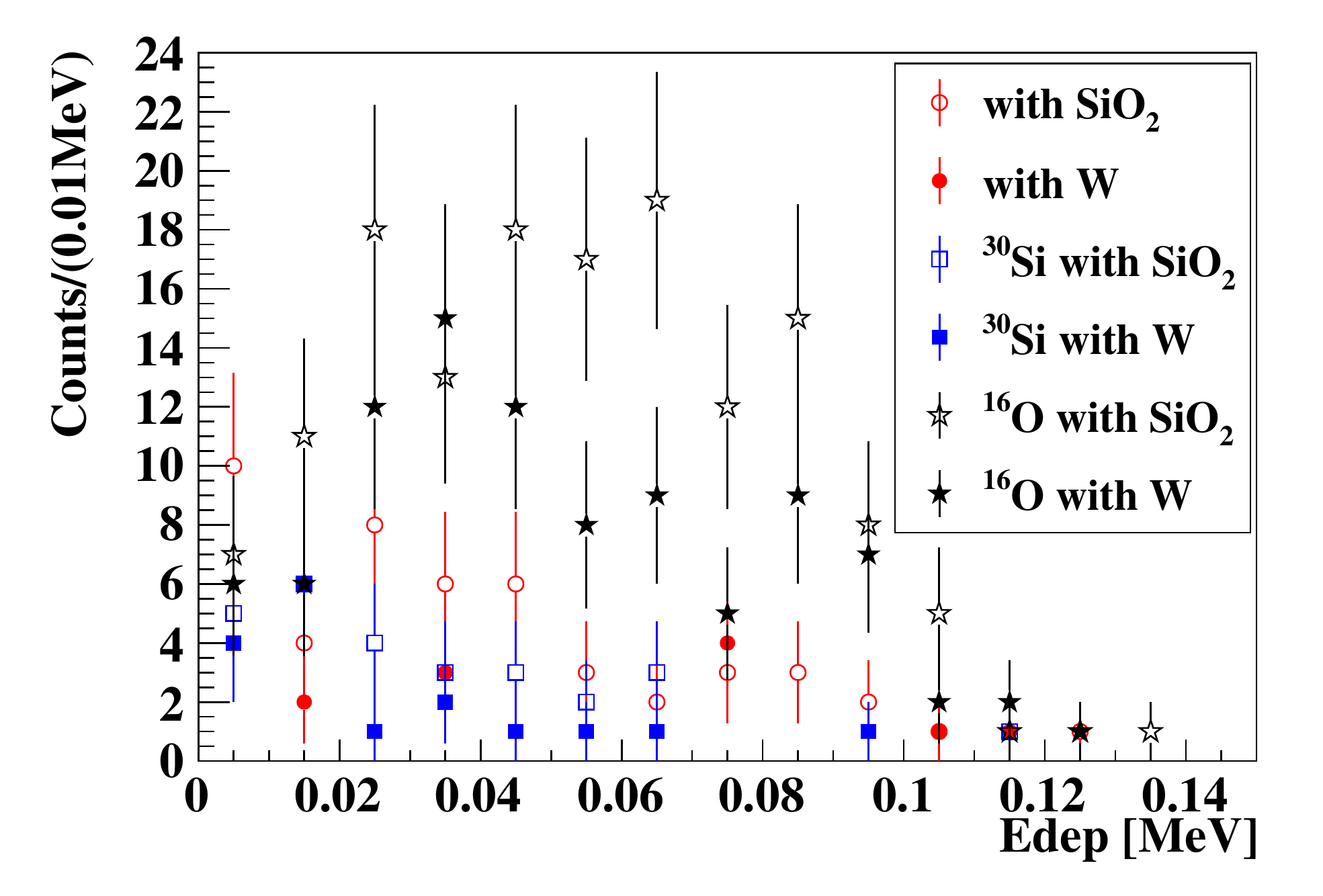}
\figcaption{\label{fig6}  Deposited energies of $^{29}\text{Si}$, $^{30}\text{Si}$ and $^{16}\text{O}$ in structures with and without tungsten layer. Total $6\times10^8$ 1-MeV neutrons are simulated. The red filled and open circles denote the deposited energy for $^{29}\text{Si}$ in structures with and without W, while the blue filled and open rectangulars for $^{30}\text{Si}$ with and without W, and black filled and open stars for $^{16}\text{O}$ with and without W.}
\end{center}

Figure~\ref{fig5} shows that in both structures, a large number of $^{28}\text{Si}$ are generated, but the energies deposited by them of each event are much less than the critical energy. The W layer is replaced by $\text{SiO}_{2}$ in the other structure, so more $^{28}\text{Si}$ are created in it, which result in more deposited energy from $^{28}\text{Si}$ in the structure with $\text{SiO}_{2}$ than that in the structure with W. The deposited energies of the rest particles for each event are also far less than the $E_{c}$ in both structures, as shown in Fig.~\ref{fig6}. According to these figures, the corresponding particles are not the reason of the increase of SEU cross section in the structure with W layer.

Figure~\ref{fig4} shows that the clear difference between the secondary particles in two structures is that many $\alpha$s are created in the structure with W layer while no $\alpha$ in the structure without it. The energy deposited by $\alpha$ of each event in the SV of the structure with W layer is shown in Fig.~\ref{fig7}.

\begin{center}
\includegraphics[width=8cm]{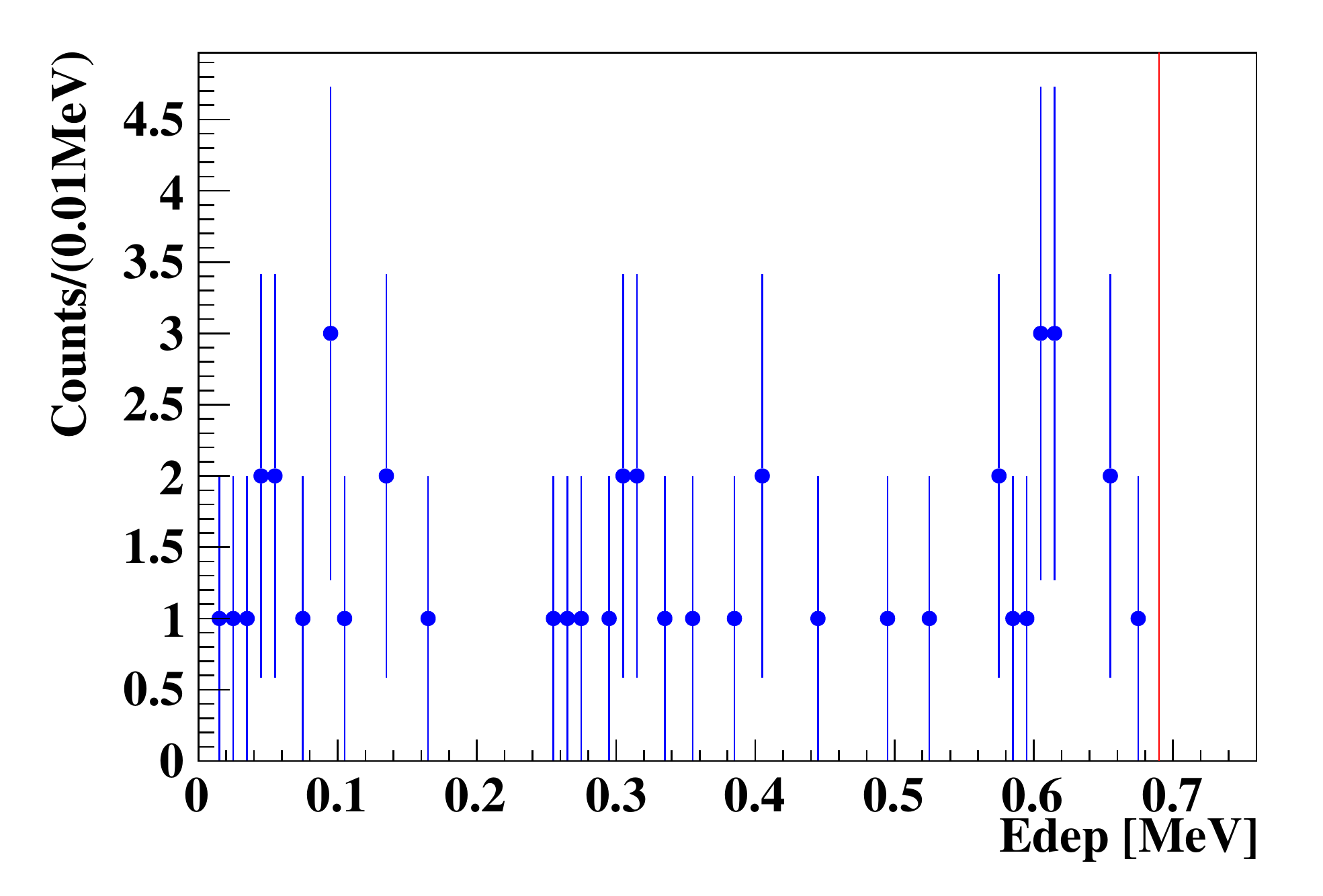}
\figcaption{\label{fig7}  Deposited energy of alpha in the SV of the structure with W layer. Total $6\times10^8$ 1-MeV neutrons are simulated. The vertical red line is for $E_{c}=0.69$\:MeV.}
\end{center}

Although there are only a small number of $\alpha$ created in the sensitive volume, most of them have much larger deposited energies than that of the other particles in the structure with W, which indicates that events with $\alpha$ created have more probability to deposit high energies and generate SEUs.

Extended studies show that when the energy of neutron is higher than {4}\:MeV, no $\alpha$ is seen in the structure without W. But more kinds of secondary particles are created and the probability  to deposit enough energy causing SEUs increases for higher energy neutron. The SEU cross sections of the two structures are similar in the relatively high energy region, as shown in Fig.~\ref{fig2}.

\section{Conclusion}

Single event upset rates induced by 1$\sim$14\:MeV neutrons in a multi-layer metallization CMOS are simulated with Geant4 in this work and the cross sections of SEU are calculated. No SEU has been seen in the structure without W when the energy of neutron is less than {4}\:MeV, while the cross sections are similar for the structures with and without W because the secondary particles other than $\alpha$ have chance to deposit enough energy to cause SEUs when the incident neutron energy is higher than 4\:MeV.

Detailed studies with 1\:MeV neutron indicate that the secondary particle $\alpha$ is created with the existence of W layer, while no $\alpha$ has been seen in the structure without W. Only events with $\alpha$ created can have chance to deposit enough energy to cause SEUs with such low energy neutron beam.
\\

\acknowledgments{We gratefully acknowledge Bo WANG, Hao QIAO, and Ying YUAN from our group, and our co-workers from Northwest Institute of Nuclear Techniques in Xi'an, for taking part of the discussions and providing suggestions.}

\end{multicols}

\vspace{15mm}

\vspace{-1mm}
\centerline{\rule{80mm}{0.1pt}}
\vspace{2mm}

\begin{multicols}{2}

\end{multicols}

\clearpage

\end{document}